\documentclass{article}

\usepackage{PRIMEarxiv}
\usepackage{tikz}
\usetikzlibrary{shapes.geometric, arrows.meta, positioning}
\usepackage[utf8]{inputenc} 
\usepackage[T1]{fontenc}    
\usepackage{hyperref}       
\usepackage{url}            
\usepackage{booktabs}       
\usepackage{amsfonts}       
\usepackage{nicefrac}       
\usepackage{microtype}      
\usepackage{lipsum}
\usepackage{fancyhdr}       
\usepackage{graphicx}       
\graphicspath{{media/}}     
\usepackage{caption}  
\usepackage{float}
\title{Predicting VBAC Outcomes from U.S. Natality Data using Deep and Classical Machine Learning Models
}

\author{
  Ananya Anand \\
  University of Illinois Urbana-Champaign \\
  Urbana, IL\\
  \texttt{ananyaa9@illinois.edu} \\
}

\begin{document}
\maketitle

\begin{abstract}
Accurately predicting the outcome of a trial of labor after cesarean (TOLAC) is essential for guiding prenatal counseling and minimizing delivery-related risks. This study presents supervised machine learning models for predicting vaginal birth after cesarean (VBAC) using 643,029 TOLAC cases from the CDC WONDER Natality dataset (2017–2023). After filtering for singleton births with one or two prior cesareans and complete data across 47 prenatal-period features, three classifiers were trained: logistic regression, XGBoost, and a multilayer perceptron (MLP). The MLP achieved the highest performance with an AUC of 0.7287, followed closely by XGBoost (AUC = 0.727), both surpassing the logistic regression baseline (AUC = 0.709). To address class imbalance, class weighting was applied to the MLP, and a custom loss function was implemented in XGBoost. Evaluation metrics included ROC curves, confusion matrices, and precision-recall analysis. Logistic regression coefficients highlighted maternal BMI, education, parity, comorbidities, and prenatal care indicators as key predictors. Overall, the results demonstrate that routinely collected, early-pregnancy variables can support scalable and moderately high-performing VBAC prediction models. These models offer potential utility in clinical decision support, particularly in settings lacking access to specialized intrapartum data.
\end{abstract}


\section{Introduction}

Cesarean sections (C-sections) account for approximately 32\% of all births in the United States, a figure that has steadily increased over the past two decades \cite{martin2023births, betran2021trends, rawashdeh2023validation}. While often medically necessary, C-sections are associated with higher risks of surgical complications—such as bleeding, infection, and prolonged maternal recovery—compared to vaginal delivery \cite{zandvakili2017maternal}. For patients with a prior C-section, two delivery options typically exist in subsequent pregnancies: elective repeat cesarean section (ERCS) or a trial of labor after cesarean (TOLAC). When successful, TOLAC results in a vaginal birth after cesarean (VBAC), which is linked to lower morbidity, shorter recovery times, and reduced health care costs \cite{habak2024vbac}. However, a failed TOLAC often necessitates emergency C-section, carrying elevated risks of uterine rupture, hemorrhage, infection, and adverse neonatal outcomes \cite{gibbs2008danforth}.

Numerous retrospective studies have identified predictors of VBAC success, including younger maternal age, prior vaginal delivery, lower pre-pregnancy body mass index (BMI), longer interpregnancy intervals, spontaneous labor onset, and favorable cervical findings such as a higher Bishop score \cite{habak2024vbac, gerhardy2022predictive, mi2021evaluation}. Conversely, maternal obesity and comorbidities like gestational diabetes and chronic hypertension are associated with decreased VBAC success rates \cite{li2019predicting, xiu2025determinants}. While these studies offer valuable clinical insight, most do not produce tools that are generalizable, scalable, or suitable for early counseling.

The Grobman calculator, developed in 2007, remains one of the most widely used tools for VBAC prediction \cite{grobman2007nomogram}. It was trained on a prospective cohort of 7,660 patients from 19 academic centers, each with a singleton, vertex fetus, one prior C-section, and gestational age beyond 36 6/7 weeks. Although the study tested several algorithms, multivariable logistic regression was selected based on lowest classification error and yielded an AUC around 0.75. One strength of the Grobman model is its use of detailed clinical variables captured at the point of care, including the indication for prior cesarean. However, its utility is limited by narrow inclusion criteria and its reliance on features only available at term or in tertiary settings, restricting its applicability in early pregnancy or in community care environments.

A recent systematic review of 57 VBAC prediction studies identified 38 unique models but found that most lacked external validation and were prone to high risk of bias \cite{black2022systematic}. While reported AUCs ranged from 0.61 to 0.95, most models were built on small institutional datasets and failed to meet TRIPOD reporting standards. Moreover, models developed closer to the time of delivery consistently outperformed those available earlier in pregnancy—highlighting a critical gap in tools for early counseling and decision-making.

To address these limitations, recent efforts have turned to machine learning (ML) techniques such as random forests, gradient boosting, and regularized regression \cite{lipschuetz2020vbac}. For instance, Meyer et al.\ \cite{meyer2022implementation} trained ML models on a dataset of 989 TOLAC cases and found that ensemble methods like XGBoost and random forest outperformed traditional logistic models in AUC-PR. However, these efforts remain constrained by limited sample sizes, narrow geographic scope, and absence of clinical integration.

Building on this prior work, the present study introduces a modern, population-scale approach for predicting VBAC outcomes using over 643,000 TOLAC deliveries from the CDC WONDER Natality dataset (2017–2023) \cite{cdcwonder2024}. This dataset spans all 50 U.S. states and territories and includes standardized prenatal variables typically available during routine care. Unlike many institutional EHR-based studies with fewer than 10,000 patients, this work leverages the largest known cohort for VBAC prediction, enabling broad geographic, demographic, and socioeconomic generalizability.

Using this nationally representative dataset, multiple machine learning models—including logistic regression, XGBoost, and a multilayer perceptron (MLP)—were trained and evaluated to assess predictive performance using features available well before delivery. Rather than solely aiming to maximize accuracy, the objective is to lay the groundwork for a scalable, clinically usable decision support tool. By leveraging early-pregnancy data and robust modeling techniques, this study moves toward a deployable system capable of supporting prenatal counseling in both high-resource hospitals and under-resourced community clinics.

\section{Materials \& Methods}
\subsection{Dataset}

The Centers for Disease Control and Prevention (CDC) WONDER Natality dataset was utilized, a nationally representative registry compiled from standardized birth certificates and local health department data collected by 57 vital statistics jurisdictions across all 50 U.S. states, the District of Columbia, and U.S. territories. The dataset includes over 25 million live births between 2017 and 2023, offering rich demographic, geographic, clinical, and obstetric detail suitable for large-scale epidemiological modeling.

Key demographic variables include maternal and paternal race, age, education level, and marital status. Geographic fields allow classification by state, urbanization level, and U.S. Census region. Clinical information includes maternal risk factors such as pre-pregnancy body mass index (BMI), tobacco use, preexisting and gestational diabetes, chronic and gestational hypertension, eclampsia, and anemia. Obstetric details encompass parity, gestational age, prenatal care utilization, infertility treatment, and prior preterm birth.

Two variables of central importance to this study are \textit{Delivery Method Expanded} and \textit{Trial of Labor Attempted (if cesarean)}, available in the expanded dataset from 2016 onward. These allow for clear classification of individuals as having attempted a trial of labor after cesarean (TOLAC) and whether the attempt resulted in a vaginal birth after cesarean (VBAC) or a repeat cesarean section. This framing enables binary classification modeling of VBAC success.

Importantly, the variables used for analysis—including maternal age, education, comorbidities, obstetric history, and early prenatal care—are typically available during routine prenatal visits. This characteristic makes the dataset highly applicable for the development of clinical decision support tools designed to guide counseling about mode of delivery during pregnancy.

\subsection{Preprocessing}

To construct the analytic cohort, natality files from 2017 to 2023 were downloaded from the CDC WONDER portal and converted into a structured CSV format using Python’s \texttt{pandas} library. An initial filter was applied to retain only singleton births where the mother had one or two prior cesarean deliveries and had attempted a TOLAC, as indicated by the relevant field. This resulted in a cohort of 738,242 delivery records.

To maintain internal validity and simplify downstream processing, records with missing or “not stated” responses for any of the 47 predictor variables were excluded. These variables spanned multiple domains—demographic, clinical, and obstetric—and included several potentially sensitive fields such as comorbidities and prenatal care history. The exclusion step resulted in a final dataset of 643,029 complete records, each suitable for machine learning and statistical analysis.

Each record was assigned a binary outcome label using the \textit{Delivery Method Expanded} variable. Deliveries resulting in successful VBAC were coded as 1, and failed TOLACs—defined as repeat cesarean deliveries—were coded as 0. This binary formulation enabled supervised classification modeling.

All continuous predictors (e.g., maternal age, BMI, gestational age) were explicitly cast as numeric types. Categorical features (e.g., race/ethnicity, marital status, census region) were encoded as factors. No imputation strategies were applied, as the large cohort size allowed for complete-case analysis without compromising statistical power. Furthermore, preterm births were retained in the sample to preserve the model’s relevance across a wide range of clinical presentations and ensure applicability to the general population.

\begin{figure}[h]
\centering
\captionsetup{justification=centering}
\includegraphics[width=0.6\textwidth]{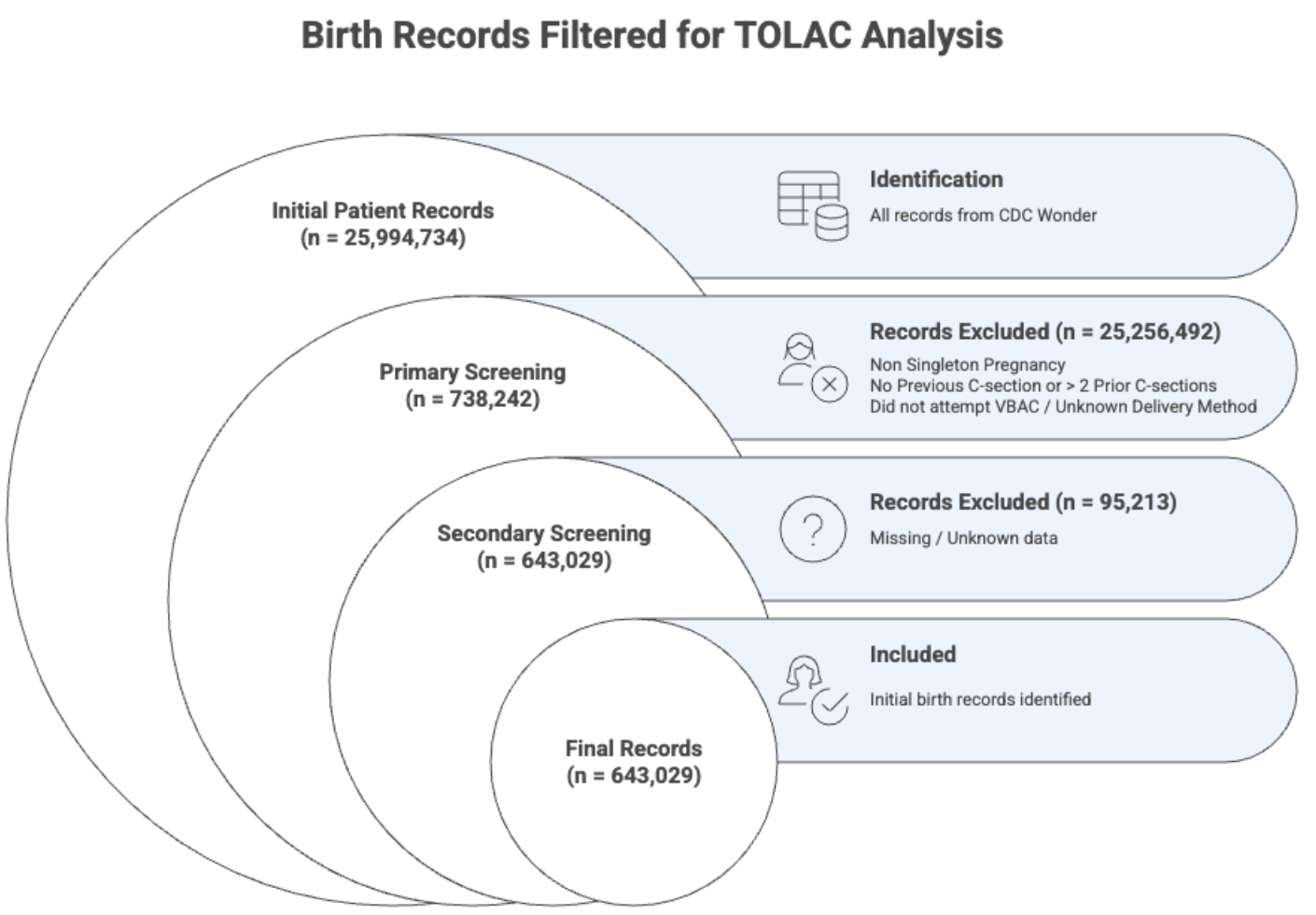}
\caption{
Overview of the cohort preprocessing pipeline.
The figure illustrates the sequence of filtering steps applied to raw natality records from CDC WONDER to derive the final analytic dataset.
}
\label{fig:processing}
\end{figure}

\subsection{Summary Statistics}

To examine differences between outcomes, the dataset was split into two groups: deliveries that resulted in a successful VBAC and those that ended in a repeat cesarean after a TOLAC attempt. Continuous variables such as maternal age, gestational age, and pre-pregnancy BMI were compared using the Mann–Whitney U test \cite{nachar2008mannwhitney}, a non-parametric method appropriate when normality cannot be assumed. A p-value below 0.05 was considered statistically significant, indicating that the distributions of the two groups differed meaningfully.

The Mann--Whitney \( U \) statistic is computed as:
\[
U = n_1 n_2 + \frac{n_1(n_1 + 1)}{2} - R_1
\]
where \( R_1 \) is the sum of the ranks for group 1 (of size \( n_1 \)), after combining and ranking all observations from both groups.

\subsection{Logistic Regression}

To establish a baseline for classification performance, a multivariable logistic regression model was implemented to predict the likelihood of successful VBAC. The dataset was randomly split into training (70\%) and testing (30\%) subsets to simulate prospective evaluation. Constant features (i.e., those with only one unique value) were dropped prior to model fitting to reduce redundancy.

The \texttt{Logit} function from the Python \texttt{statsmodels} library was used to estimate model coefficients and assess statistical significance, and a parallel logistic regression model was trained using \texttt{LogisticRegression} from \texttt{scikit-learn} for evaluation and deployment. The binary VBAC outcome served as the dependent variable. Independent variables included maternal demographics (e.g., age, race/ethnicity, education), clinical risk factors (e.g., diabetes, hypertension, pre-pregnancy BMI), obstetric history (e.g., prior cesareans, birth intervals, gestational age), and health system indicators (e.g., insurance status, place of delivery). Categorical variables were one-hot encoded, and missing values were imputed using the mode (for categorical) or median (for numerical).

Logistic regression models the probability of VBAC success using the logistic function:

\[
P(y = 1 \mid \mathbf{x}) = \frac{e^{\beta_0 + \beta_1 x_1 + \beta_2 x_2 + \dots + \beta_n x_n}}{1 + e^{\beta_0 + \beta_1 x_1 + \beta_2 x_2 + \dots + \beta_n x_n}}
\]

After training, the \texttt{predict} function was used to generate probabilities on the held-out test set. Model discrimination was assessed using the receiver operating characteristic (ROC) curve, and the area under the ROC curve (AUROC) was computed using \texttt{roc\_auc\_score} from \texttt{scikit-learn}.

To assess generalizability, 5-fold cross-validation was performed using \texttt{cross\_val\_score}. The dataset was partitioned into five equal parts, and the model was iteratively trained on four and validated on the fifth. The mean AUROC across all folds was used as a robust estimate of out-of-sample performance. Additionally, model coefficients were examined to rank predictors by absolute weight, providing interpretability into which features most influenced the predicted probability of VBAC success.

\subsection{Multilayer Perceptron}

We implemented a multilayer perceptron (MLP) using TensorFlow’s Keras API to predict VBAC outcomes from the CDC WONDER Natality dataset. The dataset was first stratified into an 80\% training and 20\% test split, preserving the class distribution of the binary outcome variable (VBAC vs. failed TOLAC). Numerical features were standardized to zero mean and unit variance using \texttt{StandardScaler}, while categorical variables were encoded via one-hot encoding. To mitigate multicollinearity, we excluded features with Pearson correlation coefficients exceeding 0.95. After preprocessing, the final feature set comprised 29 predictors spanning demographic, clinical, and obstetric factors.

The MLP model consisted of three hidden layers with 128, 64, and 32 units, respectively. Each hidden layer applied LeakyReLU activations, batch normalization, and dropout regularization (rates of 0.4, 0.3, and 0.0) along with L2 penalties ($\lambda = 0.0001$) on weights. The output layer used a sigmoid activation function to produce a probability of VBAC. Given the class imbalance (73.6\% VBAC), we computed class weights using \texttt{compute\_class\_weight} and applied them during training. The model was compiled using the Adam optimizer (learning rate = 0.0001) and binary cross-entropy loss. Training proceeded for up to 60 epochs with a batch size of 256, using early stopping (patience = 5) based on validation AUC. Model performance was evaluated on the held-out test set using AUC, precision, recall, F1-score, and a confusion matrix.

\begin{figure}[h]
\centering
\captionsetup{justification=centering}
\includegraphics[width=0.6\textwidth]{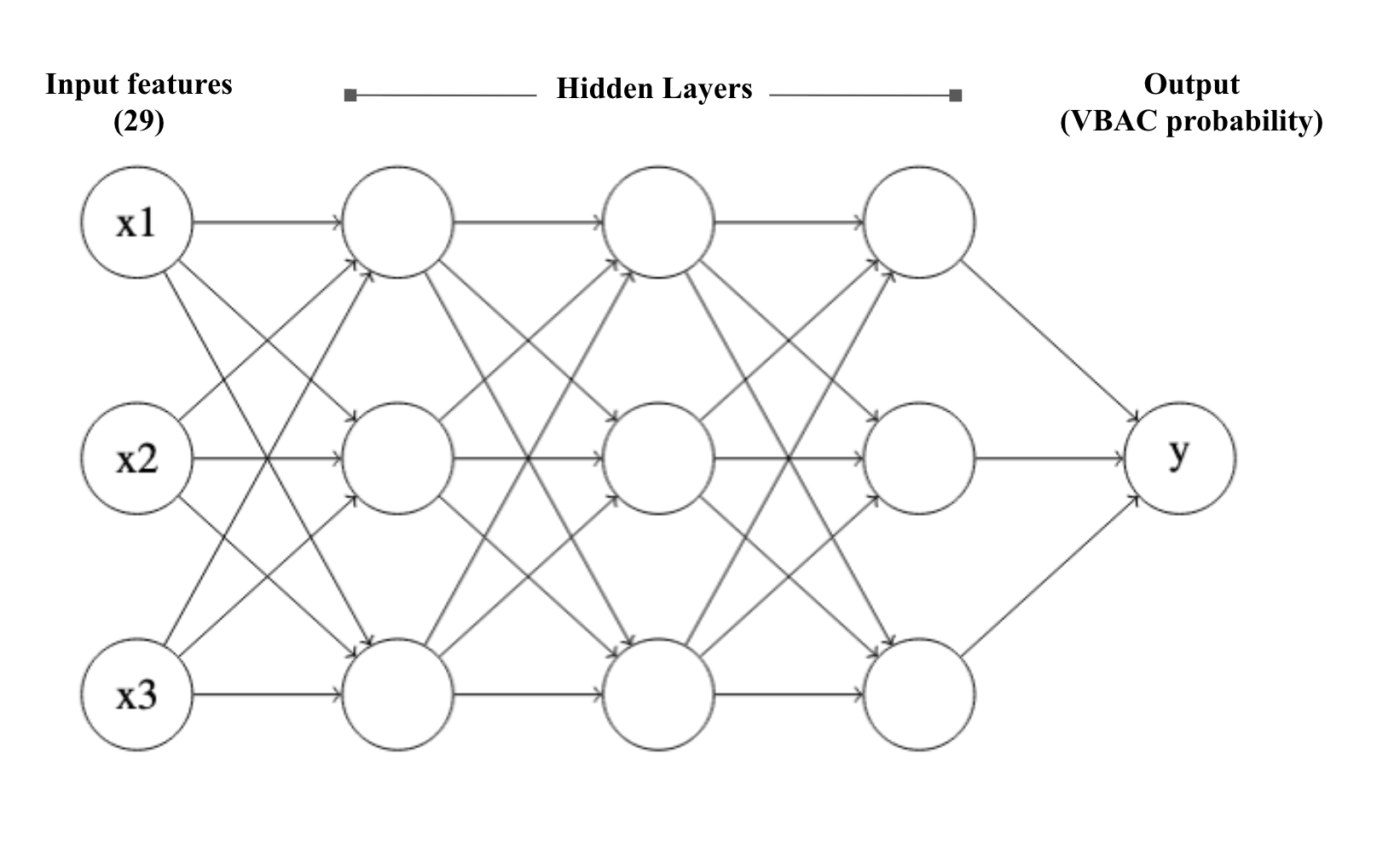}
\caption{
Structure of the MLP used for VBAC prediction. 
The model includes 3 hidden layers with 128, 64, and 32 neurons, respectively. For clarity, only a subset of nodes is shown. The final output predicts the probability of a successful VBAC.
}
\label{fig:mlp}
\end{figure}

\subsection{XGBoost Classifier}

An eXtreme Gradient Boosting (XGBoost) classifier was implemented to benchmark performance on the VBAC vs. repeat cesarean prediction task. The dataset was split into 80\% training and 20\% testing subsets using stratified sampling to maintain class distribution. Following preprocessing, highly correlated features (Pearson $r > 0.95$) were removed, leaving 29 independent predictors spanning demographic, clinical, and obstetric features.

To address class imbalance, a custom log loss objective was implemented that applied a tunable penalty (\texttt{alpha}) to misclassified instances from the minority class (VBAC) \cite{meyer2022implementation}. This weighting adjusted both the gradient and Hessian components of the XGBoost loss function dynamically during training via the low-level API. After tuning, a penalty of \texttt{alpha = 2.5} was selected to balance recall for VBACs and TOLACs, mitigating excessive bias toward the majority class.

Feature encoding was handled using one-hot encoding for categorical variables, and all numeric features were cast to appropriate types with missing values coerced where applicable. Multicollinear features exceeding a pairwise correlation threshold of 0.95 were also removed to reduce redundancy.

The model was trained using 600 boosting rounds with a learning rate of 0.01, maximum tree depth of 5, and subsample and column sampling rates of 0.9 and 0.8 respectively. Early stopping was used to prevent overfitting, halting training if validation AUC did not improve over 10 rounds. Hyperparameter tuning was guided by validation AUC on the training split. XGBoost was selected for its ability to capture non

\begin{figure}[h]
\centering
\includegraphics[width=0.65\textwidth]{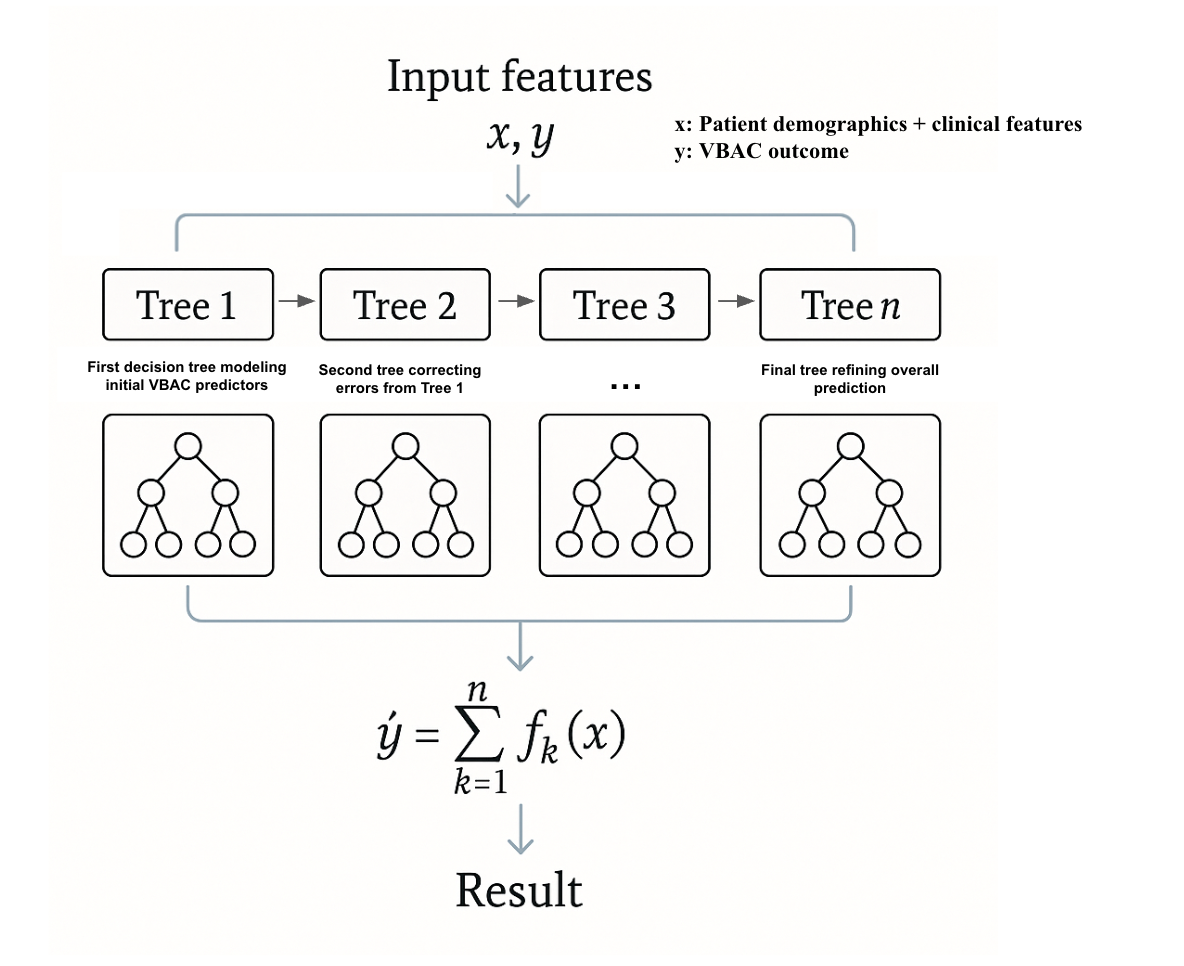}
\caption{XGBoost-based ensemble method for classification tasks.}
\label{fig:xgboost}
\end{figure}

\section{Results}
\subsection{Descriptive and Summary Statistics}

Among the 643,029 TOLAC deliveries included from 2017 to 2023, 473,016 (73.6\%) resulted in a successful VBAC, while 170,013 (26.4\%) ended in a repeat cesarean section. Table~\ref{tab:summary_stats} presents a comparison of key maternal and obstetric characteristics between the two groups, including both continuous and categorical features. This large cohort provides a comprehensive nationwide snapshot of TOLAC outcomes over a seven-year period.

Continuous variables such as maternal age, gestational age, pre-pregnancy BMI, birth weight, number of prenatal visits, and interpregnancy interval were compared using the non-parametric Mann–Whitney U test due to potential non-normality. A p-value less than 0.05 was considered statistically significant. To quantify the magnitude of observed differences, standardized effect sizes were also calculated. Categorical variables such as insurance type, delivery setting, and race/ethnicity were compared using chi-squared tests.

Maternal age was similar between groups (mean: 31 years), with no statistically or clinically meaningful difference. Gestational age was slightly lower in the VBAC group (38.6 vs 38.5 weeks), though the difference, while statistically significant ($p < 0.05$), was not clinically relevant. Notably, pre-pregnancy BMI was significantly higher in the repeat cesarean group (29.0 vs 27.2, $p < 0.0001$, effect size = 0.26), consistent with prior studies showing higher BMI as a predictor of TOLAC failure. Similar trends were observed for birth weight and prenatal visit count, though effect sizes remained modest.

\begin{table}[h]
\centering
\caption{Summary statistics comparing VBAC vs repeat cesarean deliveries}
\label{tab:summary_stats}
\begin{tabular}{lcccc}
\toprule
\textbf{Variable} & \textbf{VBAC (n = 534{,}359)} & \textbf{rCS (n = 193{,}071)} & \textbf{p-value} & \textbf{Effect Size} \\
\midrule
Age (years) & 30.95 (5.13) & 30.95 (5.27) & 0.40 & 0.00 \\
Gestational age (weeks) & 38.58 (2.67) & 38.61 (2.35) & $<$0.0001 & 0.01 \\
Birth weight (g) & 3289.21 (579.65) & 3332.37 (590.28) & $<$0.0001 & 0.07 \\
Pre-pregnancy BMI & 27.24 (6.38) & 29.03 (7.24) & $<$0.0001 & 0.26 \\
Number of prenatal visits & 10.67 (4.33) & 11.11 (4.46) & $<$0.0001 & 0.10 \\
Interval since last live birth (mo) & 46.25 (33.11) & 53.91 (38.65) & $<$0.0001 & 0.21 \\
\bottomrule
\end{tabular}
\end{table}

\subsection{Logistic Regression Performance}

\begin{figure}[h]
\centering
\includegraphics[width=0.95\textwidth]{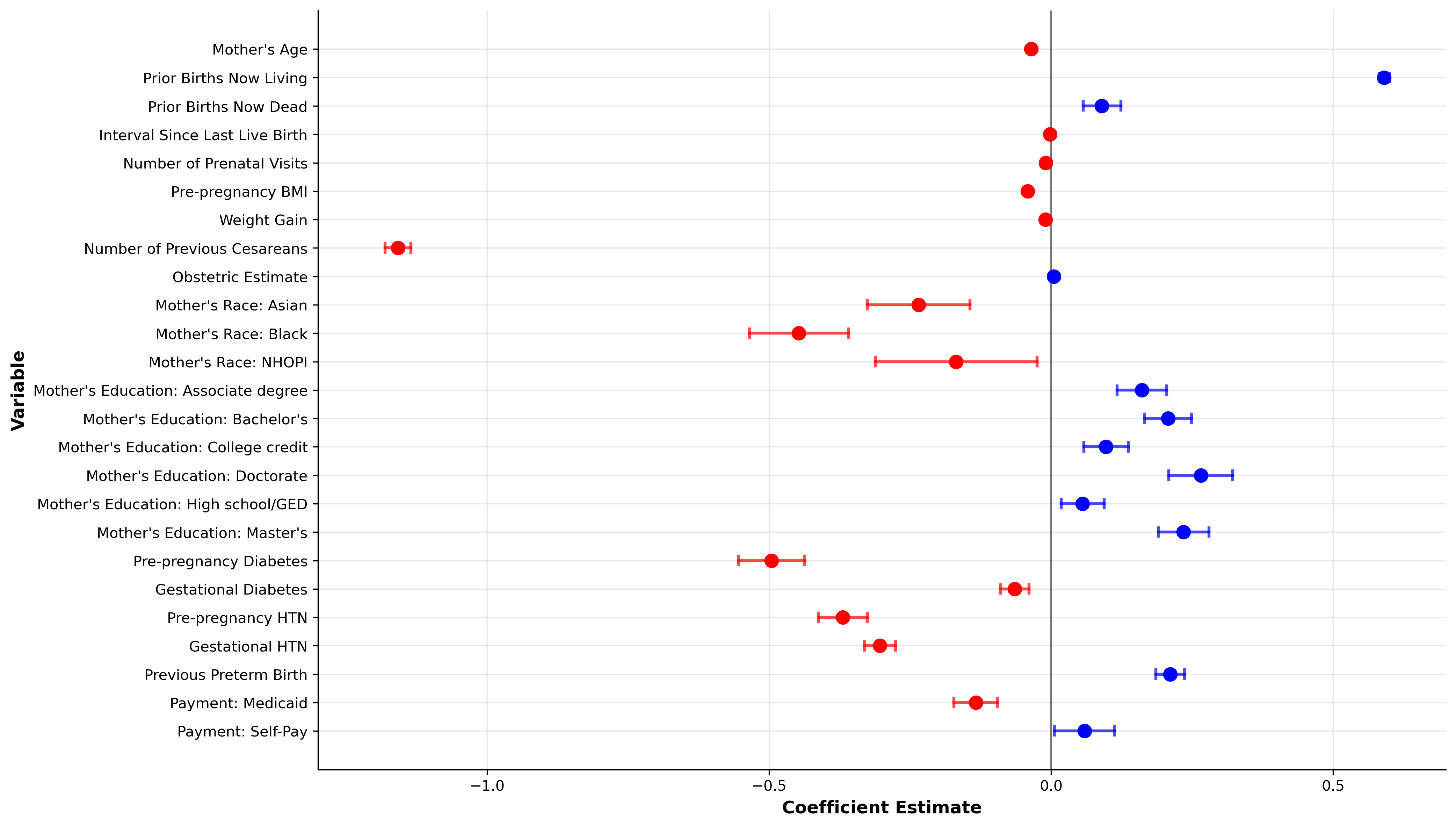}
\caption{Twenty-five of the thirty variables included in the regression were statistically significant ($p < 0.05$). Red indicates negative coefficient estimates, while blue indicates positive estimates. Error bars denote 95\% confidence intervals. BMI = Body Mass Index, NHOPI = Native Hawaiian or Other Pacific Islander, HTN = Hypertension}
\label{fig:logit_coeffs}
\end{figure}

Out of the 30 predictors included in the final logistic regression model, 25 were statistically significant ($p < 0.05$), as shown in Figure~\ref{fig:logit_coeffs}. These included well-established clinical variables such as parity, pre-pregnancy BMI, diabetes, and hypertension. Sociodemographic features, including maternal education and insurance status, also showed notable associations with VBAC success.

Variables associated with increased odds of successful VBAC included a greater number of prior live births, advanced maternal education (bachelor’s, master’s, and doctoral degrees), previous preterm birth, and higher gestational age. In contrast, decreased VBAC likelihood was associated with higher pre-pregnancy BMI, greater weight gain, more previous cesareans, and the presence of either pre-pregnancy or gestational diabetes or hypertension. Medicaid insurance coverage was negatively associated with VBAC success, while self-pay was modestly positive.

These associations are visualized in Figure~\ref{fig:logit_coeffs}, which displays model coefficients and 95\% confidence intervals. Positive predictors are shown in blue, and negative predictors in red.

The final model achieved an area under the ROC curve (AUC) of 0.709 on the held-out test set, closely approaching the reported performance of the Grobman model (AUC $\sim$ 0.75)\cite{grobman2007nomogram}. However, model performance was affected by class imbalance: while the model recalled 97\% of successful VBAC cases, recall for failed VBACs (i.e., repeat cesareans) was just 13\%. This skew underscores the need for future work incorporating class-balancing techniques such as weighted loss functions or oversampling to better capture minority-class outcomes.

\begin{figure}[H]
\centering
\includegraphics[width=0.75\textwidth]{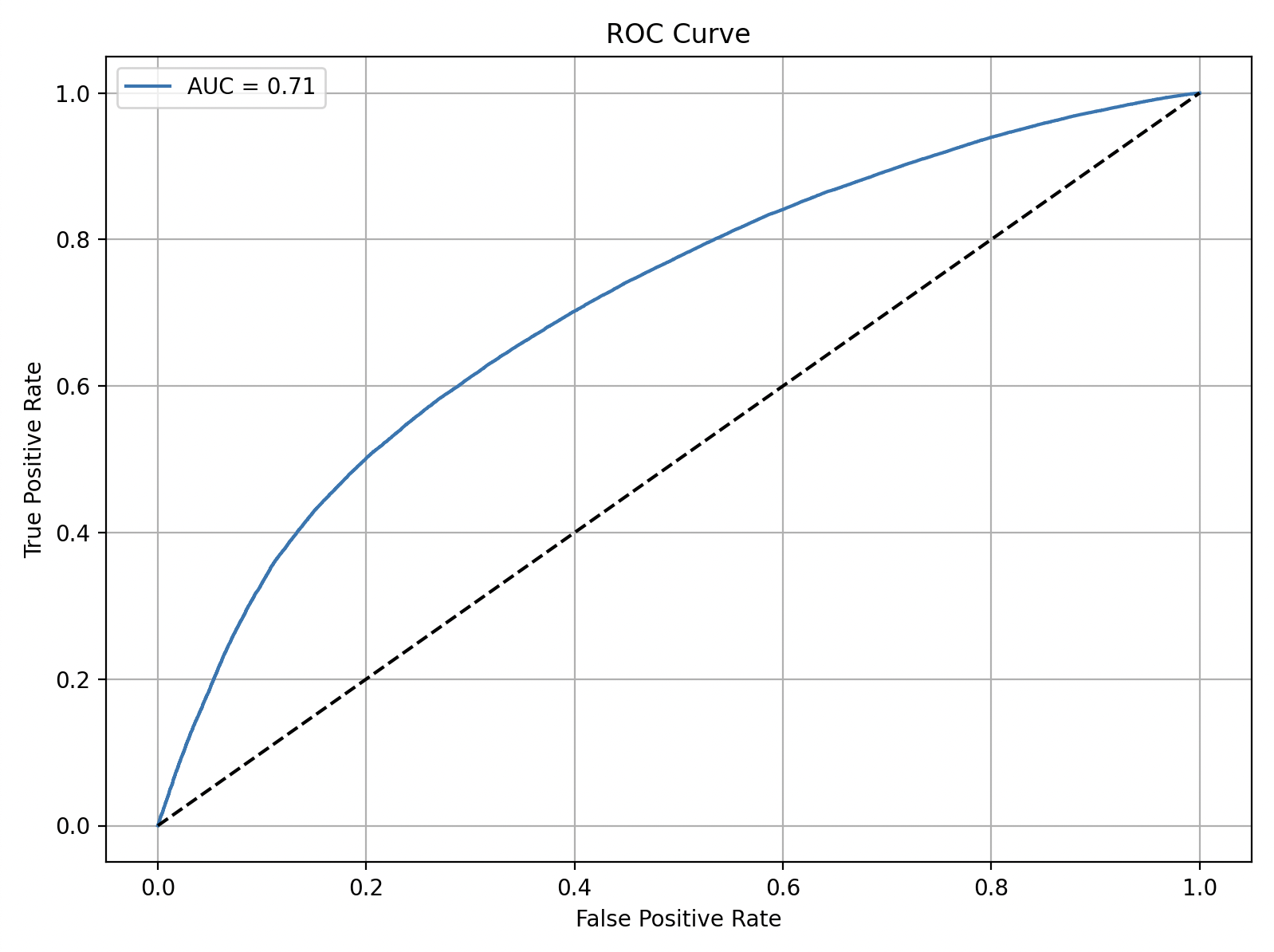}
\caption{Receiver operating characteristic (ROC) curve for the logistic regression model. The area under the curve (AUC) is 0.709.}
\label{fig:logit_roc}
\end{figure}

\subsection{MLP Performance}

The multilayer perceptron (MLP) model achieved an area under the ROC curve (AUC) of 0.7287 on the held-out test set, outperforming the logistic regression baseline (AUC = 0.709). In addition to AUC, the model attained an overall test accuracy of 63\% with a precision of 0.87 and recall of 0.59 for the majority class (VBAC), and a precision of 0.39 and recall of 0.75 for the minority class (failed TOLAC). The weighted F1-score was 0.65. These results indicate that while the model is more confident when predicting successful VBACs, it also captures a substantial proportion of failed TOLAC cases.

\begin{figure}[H]
\centering
\includegraphics[width=0.85\textwidth]{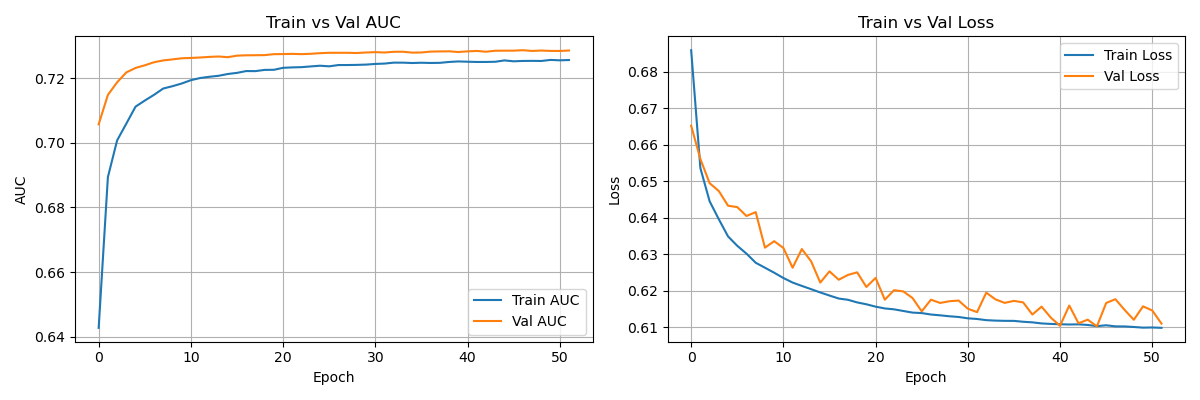}
\caption{Training and validation AUC (left) and loss (right) over 60 epochs. Early stopping occurred after epoch 51.}
\label{fig:mlp_auc_loss}
\end{figure}

As shown in Figure~\ref{fig:mlp_auc_loss}, both training and validation AUC steadily improved during the first 40 epochs before leveling off, while the corresponding loss curves declined in parallel and exhibited no significant divergence. This suggests the model learned consistent patterns without overfitting. The final model was selected based on the highest validation AUC, with early stopping halting training after 51 epochs.

\begin{figure}[H]
\centering
\includegraphics[width=0.5\textwidth]{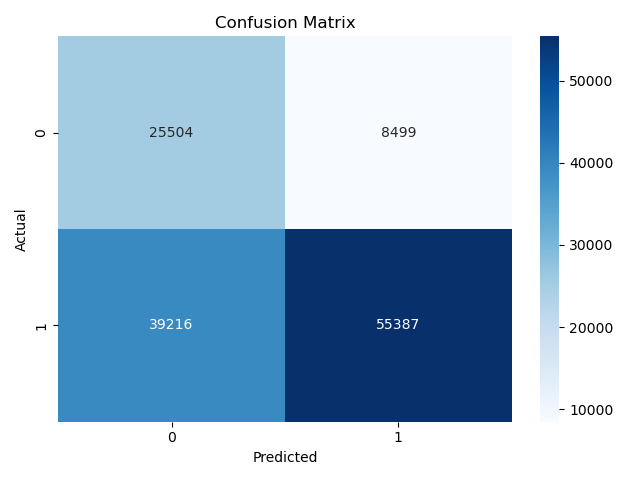}
\caption{Confusion matrix for MLP predictions on the test set. Class 0 represents failed TOLAC, while Class 1 represents successful VBAC.}
\label{fig:mlp_confmat}
\end{figure}

The confusion matrix in Figure~\ref{fig:mlp_confmat} further highlights the model's classification tendencies. Out of 94{,}603 VBAC cases, 55{,}387 were correctly predicted, while 25{,}504 of the 34{,}003 failed TOLAC cases were accurately classified. Although misclassifications remain, the model's ability to capture both positive and negative outcomes reflects reasonable clinical utility given the dataset’s limitations.

\subsection{XGBoost Results}

To evaluate a tree-based ensemble approach, an XGBoost classifier was trained with a custom loss function penalizing false negatives for the minority class (failed TOLAC). The model was trained on 80\% of the dataset and evaluated on the remaining 20\% test set. Hyperparameters such as learning rate ($\eta=0.01$), tree depth (5), and regularization strength were selected empirically to prevent overfitting.

The final model achieved a test set ROC-AUC of 0.727 and a PR-AUC of 0.879, indicating strong discrimination and precision-recall tradeoffs despite the class imbalance. The optimal threshold for classification, selected based on maximum F1 score, was 0.10.

\begin{figure}[h]
    \centering
    \includegraphics[width=0.6\textwidth]{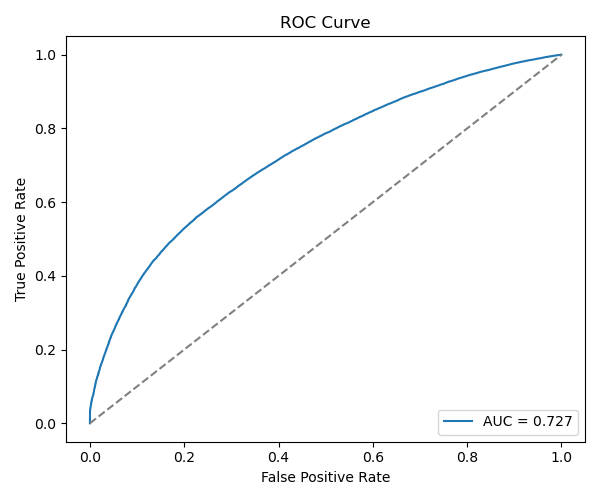}
    \caption{ROC curve for the XGBoost model on the test set.}
    \label{fig:xgb_roc}
\end{figure}

As shown in Figure~\ref{fig:xgb_roc}, the ROC curve indicates better-than-random performance with a substantial area under the curve. However, the relatively low recall for VBACs at higher thresholds motivated the use of a lower cutoff to improve sensitivity.

\begin{figure}[h]
    \centering
    \includegraphics[width=0.75\textwidth]{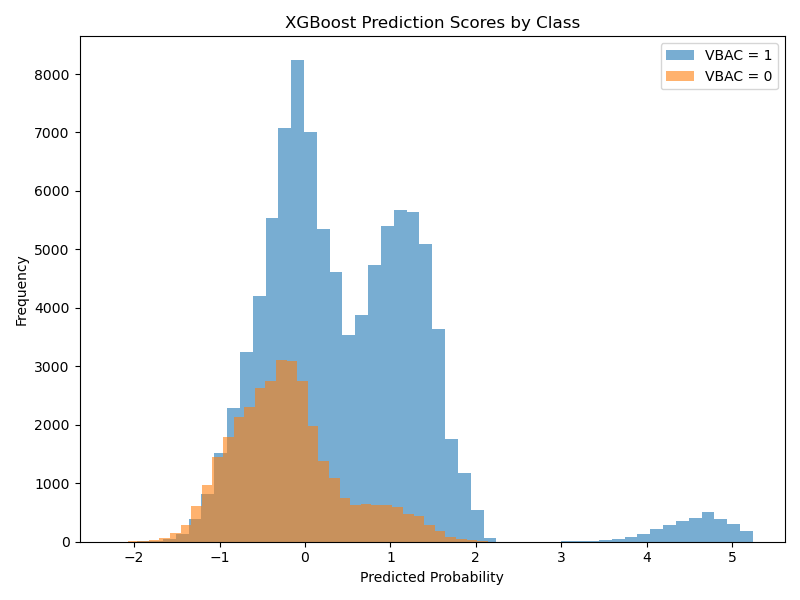}
    \caption{Distribution of predicted XGBoost scores by class.}
    \label{fig:xgb_histogram}
\end{figure}

Figure~\ref{fig:xgb_histogram} shows the distribution of predicted scores by true class. While there is notable overlap, the separation is sufficient to enable threshold-based discrimination.

Applying the 0.10 threshold, the model achieved a VBAC precision of 0.86 and recall of 0.59, while recall for C-section cases reached 0.74. Overall accuracy was 63\%, with a weighted F1-score of 0.65.

\begin{figure}[h]
    \centering
    \includegraphics[width=0.6\textwidth]{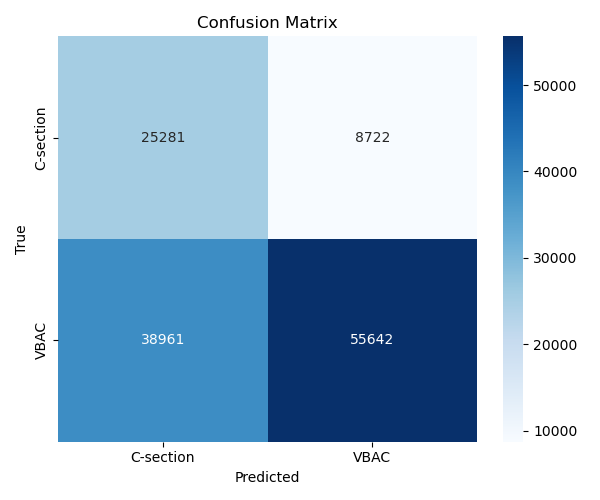}
    \caption{Confusion matrix of the XGBoost model at optimal threshold (0.10).}
    \label{fig:xgb_cm}
\end{figure}

Figure~\ref{fig:xgb_cm} displays the confusion matrix, highlighting that while many VBACs were correctly identified (TP = 35,297), a large number were misclassified as C-sections (FN = 59,306). This tradeoff reflects the conservative tuning of the model to improve negative predictive value (NPV = 0.39) and reduce missed C-section cases. In clinical settings, such tuning may be warranted to prioritize safety over specificity in VBAC recommendations.

\section{Discussion}

Across all three models—logistic regression, XGBoost, and multilayer perceptron (MLP)—consistent predictive performance was observed, with AUC values ranging from 0.709 to 0.7287. While the MLP achieved the highest AUC overall, the margins of improvement were modest—a trend supported by Wollmann et al.\ \cite{lindblad2020predicting} in a Swedish-centered study as well as Thagaard et al.\ \cite{thagaard2024prediction}. These findings suggest that while non-linear models can better capture complex interactions among variables, the available features from the CDC WONDER dataset may place an upper bound on predictive performance. In other words, the models likely approach the ceiling of what is learnable from birth certificate data alone.

One notable observation is the trade-off between model complexity and interpretability. Logistic regression provided the most transparent insights, highlighting strong negative associations between VBAC success and variables such as higher BMI, diabetes, and hypertension—factors well-documented in prior literature \cite{mi2021evaluation, wu2019factors}. In contrast, the MLP and XGBoost models showed improved recall for failed TOLAC cases, especially when class-weighting and custom loss penalties were applied. These results suggest that ensemble and neural methods are more effective at learning subtle, non-linear patterns, especially for the minority class. However, the interpretability of these models remains a barrier to direct clinical translation.

The relatively poor recall for failed TOLACs in the logistic regression model (13\%) highlights the limitations of linear approaches in class-imbalanced settings. In contrast, the XGBoost model achieved a recall of 74\% for failed TOLACs when using a lower decision threshold (0.10), but at the expense of increased false positives for VBAC. This reflects a critical trade-off in clinical deployment: conservative thresholds can reduce false reassurance but may also deter candidates who would have succeeded with TOLAC. Despite these improvements, the models still underperform compared to reported benchmarks in smaller clinical datasets—some achieving AUCs as high as 0.85–0.90 \cite{black2022systematic}. This discrepancy may be attributable in part to the coarseness and limited clinical depth of the CDC WONDER dataset, which lacks real-time labor metrics such as cervical dilation, fetal heart rate patterns, or provider-level variables that likely influence delivery outcomes.

Several limitations warrant further investigation. Although the CDC WONDER dataset provides unmatched population-level scale, it lacks intrapartum features such as cervical exam findings, fetal monitoring data, and provider characteristics that are routinely available in hospital electronic health records (EHRs). To address this, the trained models could be integrated into prenatal care workflows through an interactive web-based tool or EHR-integrated module that takes in early pregnancy variables—available at booking or early prenatal visits—and provides clinicians with a real-time VBAC probability estimate. Such a tool would allow exploration of "what-if" scenarios (e.g., changes in BMI or comorbidity profile), while also supporting shared decision-making conversations. Future work will focus on validating the model in hospital EHRs and refining the tool’s interface through usability testing with OB/GYN stakeholders.

\section{Conclusion}

This study developed and evaluated predictive models for vaginal birth after cesarean (VBAC) using over 643,000 TOLAC cases from the nationally representative CDC WONDER Natality dataset. The findings indicate that modern machine learning methods—specifically XGBoost and multilayer perceptrons—offer slight but consistent improvements over traditional logistic regression, with the MLP achieving the highest test AUC of 0.7287. These results underscore the potential for scalable, population-level tools to support individualized prenatal counseling using routinely collected data.

Notably, models built exclusively on prenatal-period features—those typically available well before labor—demonstrated reasonable predictive performance, even in the absence of real-time intrapartum variables. This supports the feasibility of early decision-making around TOLAC candidacy, particularly in contexts lacking access to tertiary care or advanced diagnostics.

Despite these advancements, challenges remain. Predictive performance for failed TOLAC cases remains limited, especially under conditions of class imbalance. Addressing this issue may require targeted algorithmic strategies, as well as the incorporation of richer clinical variables from electronic health records. Furthermore, clinical integration demands a focus on interpretability and usability, necessitating continued work on explainable AI, user interface design, and deployment in diverse care settings.

This work establishes a foundation for a modern, data-driven VBAC prediction tool that is both scalable and generalizable, with the potential to inform shared decision-making at a national scale. Future efforts should focus on embedding these models into clinical workflows as user-facing decision support systems, incorporating feedback loops and validation within OB/GYN EHR environments to enable early, personalized counseling around mode of delivery.

\section*{Acknowledgements}

I would like to thank Anisha Mittal and Peyton Paschell for being the most supportive mentors and for making this project possible through their encouragement and guidance. I am also grateful to the CDC for providing public access to the WONDER Natality dataset, which was foundational to this study.

Finally, I sincerely appreciate the peers who offered thoughtful feedback and comments that greatly improved the quality and clarity of this work.

\bibliographystyle{unsrt}  
\bibliography{references}

\end{document}